\documentclass[graybox]{svmult}

\usepackage[utf8]{inputenc}

\usepackage{type1cm}        
\usepackage{makeidx}         
\usepackage{graphicx}        
\usepackage{multicol}        
\usepackage[bottom]{footmisc}

\usepackage{natbib}
\newcommand{\enzo}{\it{\small ENZO}}

\usepackage{newtxtext}       %
\usepackage{newtxmath}       

\makeindex             

\begin{document}

\title*{The Complexity and Information Content of Simulated Universes}
\author{Franco Vazza}
\institute{Franco Vazza \at  Dipartimento di Fisica e Astronomia, Universit\'{a} di Bologna, Via Gobetti 92/3, 40121, Bologna, Italy;  Hamburger Sternwarte, Gojenbergsweg 112, 21029 Hamburg, Germany;
Istituto di Radio Astronomia, INAF, Via Gobetti 101, 40121 Bologna, Italy. \email{franco.vazza2@unibo.it}}
%
%
\maketitle

\abstract{
The emergence of a complex, large-scale organisation of cosmic matter into the Cosmic Web  is a beautiful exemplification of how complexity can be produced by simple initial conditions and  simple physical laws.  In the epoch of Big Data in astrophysics, connecting the stunning variety of multi-messenger observations to the complex interplay of fundamental physical processes is an open challenge.  In this contribution, I discuss a few relevant applications of 
Information Theory to the task  of objectively measuring the complexity of modern numerical simulations of the Universe. When applied to cosmological simulations, complexity analysis makes it possible to measure the total  information necessary to model the cosmic web. It also allow us to  monitor which physical processes are mostly responsible for the emergence of complex dynamical behaviour across cosmic epochs and environments, and possibly to improve mesh refinement strategies in the future.}
 
\newpage

\section{Introduction}
\label{sec:1}
{\it "I think the next century will be the century of complexity." Stephen Hawking, Complexity Digest 2001/10, 5 March 2001 \\}

\bigskip

{\it “Don’t go on multiplying the mysteries,’ Unwin said. 'They should be kept simple. Bear in mind Poe’s purloined letter, bear in mind Zangwill’s locked room.’ 'Or made complex,’ replied Dunraven. 'Bear in mind the universe.” 
(Jorge Luis Borges, The Aleph and Other Stories)}\\

The description of physical processes in Nature often calls for the concept of "complexity", as the reason why achieving a satisfactory and quantitative description of a particular phenomenon is challenging or just impossible. Astrophysics and Cosmology make no exception.
"Complexity" is generally regarded as a difficulty inherent to the many degrees of freedom present in a system, or to the difficulty to compute its evolution, e.g. by direct integration of differential equations.  

In the present epoch of "Big Data", driven by ever-growing multi-wavelength observing facilities, 
a continuous struggle for astrophysicists is the one of connecting the stunning variety of observations to the complex dynamics behind their origin, with the final goal of establishing the underlying physical processes and initial conditions.

This challenge requires the development of new analysis techniques, derived from fields outside of standard astrophysics, and scalable up to large datasets. 
 
 It is of outstanding importance for astrophysics also to explore radically different approaches, capable to identify  information-rich pattern in real or simulated data sets, without available pre-labeled training sets.  

Information Theory (IT) is a powerful and multidisciplinary field of investigation, which enables a mathematical representation of the conditions and parameters affecting the processing and the  transmission of information across physical systems \citep[e.g.][]{Glattfelder2019}.  According to IT, all physical systems - the entire Universe included - can be regarded as an information-processing device, which computes its evolution based on a software made of physical laws. Thanks to IT,  the complexity of a process becomes a rigorous concept, which can be measured and compared, also between different fields of research \citep[e.g.][]{prokopenko2009information}. 
 In IT, not all systems whose evolution is complicated to compute or to predict are truly {\it complex} in a physical sense. For example, a purely random process does not allow a precise prediction of its future state, yet its future evolution be trivial to compute in a statistical sense.  On the other hand, a truly complex phenomenon demands a significant amount of information in order to predict its future evolution, even in a statistical sense. \\
 
 Our representation of the Universe, based on the ever-growing collection of multi-wavelength telescope observations gives us the image of an arguably very complex hierarchy of processes, spanning an outstanding range of spatial and temporal scales, leaving us with plenty of unanswered questions. 

How and when did the cosmic structure come into shape? 
How did galaxies and the matter connected to them form and shape the Universe we can observe with telescopes? 
Which processes are fundamental to explain the observed richness of cosmic structures, and which ones can be neglected to the first degree of approximation? 

The quantitative answer to such questions is a challenge in which analytical, semi-analytical or numerical methods are struggling since almost a century.
Decades of research suggests that large-scale cosmic structures emerged from a hierarchy of interconnected processes, in which several mechanisms (e.g. the expansion of the space-time, gravity, hydrodynamics, radiative and chemical gas processes, etc.) have coupled in a non linear way.  Cosmic matter self-organised  across an enormous range of scales, transitioning from the  smoothest and simplest possible initial condition (a nearly scale-invariant background of matter fluctuations, $\delta \rho/\rho \leq 10^{-5}$, embedded in an expanding space-time, where $\rho$ is the matter density) into a spectacular hierarchy of clustered sources, with a final density contrast of $\delta \rho /\rho \geq 10^4-10^5$ 
\citep[e.g.][]{1993ppc..book.....P, 1985ApJS...57..241E,KA99.3,2005Natur.435..629S,2014Natur.509..177V}. 

This paradigm perfectly fits into the standard definition of how a complex system arises{\footnote{See https://complexityexplained.github.io for a recent public repository of resources and visualization tools to explore complexity in physics.}}: complexity  is often found to emerge from a minimal set of (seemingly simple) initial conditions and physical laws. 
Moreover, the observable clustering properties of the Universe cannot be  predicted just based on its main build blocks alone (e.g. galaxies or dark matter halos), but emerged from the interplay between many components and many scales of interaction. 

Numerical simulations are the perfect tool to study how a large number of discrete elements can produce complex collective behaviours through their network of interactions.

\begin{question}{Which aspects of cosmic structure formation can benefit from complexity analysis?}

In the digital representation of our Universe by supercomputers,  the emergence of complex dynamics out of simple initial conditions is made manifest by the fact that a single random string of a few digits, combined with a source code that can be stored in a few $10^2$ Mb (linked to more external numerical libraries and compilers) can produce extremely rich and structured systems, which require tens or hundreds of Terabytes of disk space to be stored. 
For example, the widely used Smoother-Particle-Hydrodynamics (SPH) cosmological code {\small GADGET-2} (http://www.mpa-garching.mpg.de/gadget/)  has a compressed size of  $\sim 200 \rm kb$, but the largest Magneticum simulations (magneticum.org) need $\sim 20$ Tb to store each of snapshots. 
The moving-mesh code {\small AREPO} (http://arepo-code.org) has a compressed size of $\sim 12$ Mb, but the final raw dataset produced by the Illustris-1 simulation is of $\sim 200$ Tb .
The  latest version of the adaptive mesh refinement code {\small ENZO} used in this work (https://code.google.com/p/enzo/) has a compressed size of $\sim 2.1 \rm Mb$, but the outputs of the latest Renaissance runs (https://rensimlab.github.io) need $\sim 100$ Tb of disk space.\\

Alongside the growth of the size of the observational dataset planned for future surveys of the sky, cosmological simulations continue to produce larger and larger simulations, with the ambitious goal of representing a big fraction of the observable Universe  with a  high enough resolution to properly resolve galaxy formation \citep[e.g.][for a recent review]{2019arXiv190907976V}. 
The largest cosmological simulations have indeed stepped into the regime of evolving $O(10^{11})$ resolution elements, often storing 3-dimensional properties of gas and dark matter dynamics, chemical composition, star-forming properties and magnetic fields attached to each of them. 
Data mining in such colossal datasets represents  a challenge, for which "standard" analysis methods are continuously being deployed and optimised. Also, the preliminary choice of which dynamical scales and volumes are essential to simulate is often a non-trivial one before starting every extensive simulation campaign. 

In this respect, the development of efficient and  objective tools to measure the emergence of complexity in any numerical model enables 
simulators to assess which spatial scales are responsible for complex phenomena observed by telescopes. This allows simulators to deduce which are the relevant scales for minimal working representation of the cosmic dynamics, which is also crucial to match the 
extensive sampling of cosmic volumes and redshift space that future multi-band surveys of the sky will deliver (e.g. from Euclid to the Square Kilometer Array, e.g. \citealt{2018Galax...6..120F}). 

 While there have been valuable attempts to define and study complexity in several physics topics (e.g. climate data analysis, \citealt[e.g.][]{hoffman2011data}, cellular automata \citealt[][]{1984PhyD...10....1W}, limnology \citealt[][]{2013arXiv1304.1842F}, epidemiology \citealt[][]{escidoc:2219209}, compact stars \citealt[][]{deAvellar20121085}, and many more), the application of Information Theory to the structure formation paradigm has just begun \citep[][]{va17info,va19info}. 

As I will discuss in this contribution (Sec. 3.1) complexity analysis can also be applied at run-time and be instrumented to cosmological codes, to enable them to refine numerical simulations on the fly, wherever complex dynamical patterns are formed. 

Moreover, the association of complexity with ubiquitous matter accretion towards cosmic structures is very significant, and the same accretion phenomena are also believed to power diffuse non-thermal emissions, which are one of the most critical scientific driver of existing and future radio surveys \citep[][]{brown11,vern17,2019SSRv..215...16V}. The fact that the emergence of non-thermal phenomena and complex evolutionary patterns are tightly associated (see, e.g. Sec.3.2) means that any progress in the understanding of how complexity has emerged in the Universe will concern the same environment that the largest astronomical enterprise of this decade, i.e. the Square Kilometer Array \citep[e.g.][]{2004ApJ...617..281K,va15ska,va15radio}, will be mostly devoted to investigate. \\ 
\end{question}

\section{Information and complexity: an overview}

In this Section I give a compact overview of the key concepts from  Information Theory, whose origin is commonly fixed to the seminal work by \citet{1949IEEEP..37...10S} and ~\citet[][]{1949mtc..book.....S}, which were concerned on  signal processing in communications. 
The following methods have been applied in \citet{va17info} and later on in \citet{va19info} to cosmological simulations. 
 For more general details on the methods, I refer interested readers to the excellent review by \citet[][]{prokopenko2009information}. \\

\subsection{Shannon's information entropy}

Information Theory posits that the information content of the outcome a probabilistic process, $x$ (with probability $P (x)$) can be defined as 

\begin{equation}
    \log_2\frac{1}{P(x)}=-\log_2[P(x)],
    \label{shannon}
\end{equation}
measured in  {\it bits} \citep{1949mtc..book.....S}. The latter is known as {\it information entropy}, and it measures the 
degree of randomness contained in the process. 
Therefore, crucial for any attempt to quantify information and complexity is the consideration that any physical phenomenon  can be regarded as an information processing device, whose evolution produces a sequence of outputs (e.g. energy states), which can be analysed through symbolic analysis. 

The latter approach also implies that a process with many different possible outcomes has high entropy, and that this measure is a proper quantification of how much choice is involved in the realisation of a specific event (i.e. a specific sequence of symbols). Following from this,  the complexity of a system equals to the amount of information needed to fully describe it. The  strict connection between how unlikely is for a specific sequence of events/symbols to occur and the amount of information necessary to describe such sequence is key to any modern description of complexity.  

\subsection{The algorithmic complexity}
The minimal information needed to perfectly (i.e. without any loss of information) describe a phenomenon or system is measured by the {\it algorithmic complexity} \citep[e.g.][]{kolm,1995chao.dyn..9014C}. In numerical simulations, this is roughly connected to the disk memory necessary to store every single digit produced by the simulation itself, or by the entire source code used for the simulation as well as its initial conditions.  
 Such representation of complexity poses some practical problems, which are best explained by thinking to it as to a compression problem{\footnote{http://www.ics.uci.edu/\~dan/pubs/DataCompression.html}}: a simple periodic object is characterised by a very little algorithmic complexity as it can be very significantly compressed because the source code to generate a periodic system can be very short (e.g. a cosinusoidal function).
However, the algorithmic complexity for a purely random sequence 
of data can be enormous (i.e. of the order of the sequence itself), due to the lack of internal structure and to the impossibility of further compressing it. 

Therefore, this definition of complexity does not entirely capture the basic physical intuition of natural or artificial phenomena: for example, an arbitrarily long sequence of rand digits has a higher Kolomogorov complexity than the sequence of velocity fluctuations in a turbulent fluid, of the sequence of orbits of planets in the Solar system, or 
of a symphony. Our physical intuition views instead all of the latter phenomena as more "information rich" than any purely random sequence of numbers. 
For this reason alternative approaches to the measure of complexity in natural systems have been developed. 

\subsection{The statistical complexity}
\label{stat_complex}
More relevant from the physical perspective is the quantification
of how much information is needed to statistically describe the evolution of a system:  
 this is given by the {\it statistical complexity} \citep[e.g.][]{adami}. 
 The statistical complexity quantifies the similarity between different realisations of the same process (e.g. starting from different randomly drawn initial conditions) as well as how likely it is that different outputs are drawn from the same process.  It also measures the amount of  information needed to produce a sequence of symbols which is statistically similar  to the original sequence of symbols under study. 
 
At the practical level, the statistical complexity, $C_\mu$, is measured by partitioning the internal states of a system into $N_{\rm bin}$ discrete levels (e.g. internal energy levels), followed by calculating the conditional probability, $P(E_2|E_1)$, that elements in the system evolve from level $E_1$
into level $E_2$ going from epoch $t$ to epoch $t+\Delta t$. 

The evolution of each element in the simulation is tracked over time, searching for patterns.  If an element always gives the same output, its evolution is simple to prescribe, and the statistical complexity is small. On the other hand,  elements which require a  large amount of information in order to prescribe their evolution are complex.\\

The probability distribution function $P(E_2|E_1)$  of  possible transitions between the states $E_2$ and $E_1$ can be traced a-posteriori in the data stream, by building a matrix of all recorded transitions across the simulation's elements. 
From the entire matrix of transition probabilities  it is thus possible to calculate the invariant probability distribution $P(E)$, over the entire sequence of causal states in the system'history, as well as its 
associated information content as the Shannon entropy of all transitions:

\begin{equation}
C_\mu = - \sum_{E_{i}} P_i(E_2|E_1) \log_2 P_i(E_2|E_1) \rm [bits].
\label{eq:complex}
\end{equation}

\noindent where the summation is performed over all computing elements in the simulation at a given time. Each single computing element  has thus a complexity given by  $C_{\mu,i}=-\rm log_2 P_i(E_2|E_1)$. 

In general, there is no unique way of partitioning the internal energy levels of a specific simulation. The exact partitioning strategy of the system is the result of a compromise between the need of keeping the computing resources under control (as the computation of the statistical complexity scales as $\propto N_{\rm bin}^2$ (where $N_{\rm bin}$ is the number of energy bins), and the need of 
 resolving all relevant energy transitions between close timesteps.


\subsection{The block entropy and the entropy gain}
\label{subsubsec:block_entropy}

The probability of observing a specific sequence of symbols, generating a string of length $L$, is
\begin{equation}
H(L) = - \sum_{x^L \in X^L} P(x^L) \log P(x^L) \rm [bits]
\label{eq:block_entropy}
\end{equation}
where $X^L$ contains all possible sequences of symbols with length $L$ in the datastream, and is called 
{\it block entropy} \citep[e.g.][]{Larson20111592}. 

The block entropy is a monotonically increasing function of $L$ \citep[e.g.][]{feldman1997bii,Crutchfield03}, and 
the increase of $H(L)$ is called {\it entropy gain}:

\begin{equation}
h_\mu(L)=H(L)-H(L-1). 
\end{equation}

The entropy gain is a good proxy for the intrinsic randomness in a sequence of symbols, because it measures  the information-carrying capacity of a string of $L$ symbols, and it quantifies the internal level of correlation.
In the limit of large $L$, such metric converges to $H(L)/L$: 
\begin{equation}
h_\mu=\lim_{L\rightarrow \infty} h_\mu(L)=\lim_{L\rightarrow \infty}  H(L)/L,
\end{equation}

which is usually called {\it source entropy rate}. 

\subsection{The excess entropy and the efficiency of prediction}
\label{subsubsec:eff}

The information due to correlation over larger blocks (i.e. due to the intrinsic redundance of the source of symbols) is the {\it excess entropy}: 

\begin{equation}
E= \sum_{L=1\rightarrow \infty} [h_\mu (L)-h_\mu].
\end{equation}

The excess entropy can be interpreted as the apparent memory of structure in a source of $L$ symbols \citep[e.g.][]{2004PhRvL..93n9902S}. $E$ can be further simplified into a finite partial-sum  for a length $L$:

\begin{equation}
E(L)=H(L) - L \cdot h_\mu (L).
\end{equation}

Systems with a large dynamical range allow the observer to describe them on a variety of scales. 
 The {\it efficiency of prediction}, $e$, quantifies  the scale at which making future predictions of the system  gives the best "emergent" and information-rich view \citep[][]{2004PhRvL..93n9902S}: 
 
\begin{equation}
e=\frac{E}{C_\mu},
\end{equation}

\noindent i.e. the ratio between the excess entropy and the statistical complexity. 
The spatial scale at which $e$ is maximum allows an observer to make the most informative predictions of the future evolution of the system. Indeed, while the excess entropy $E$ uses the past evolution of the system to predict its future evolution, while the complexity $C_\mu$ gives the amount of information necessary to statistically reproduce its behaviour. 
Hence  $e=E/C_\mu$ can be regarded as a proxy for ``how much can we predict'' compared to ``how much difficult it is for us to predict'' about the evolution of a system \citep[][]{prokopenko2009information}.

\section{Results}

\label{sec:2}

 In the following Sections I will give an overview of the main results 
 concerning the study of the complexity of large-scale structures in the cosmic web using numerical grid simulations and various proxies for complexity, extending first results presented in \citet{va17info} and \citet{va19info}. 
 Section 3.1 focuses on the analysis at high-resolution of a massive galaxy cluster while  Sec.3.2 presents a more global view of the cosmic web. 
All numerical simulations used in this work are Eulerian (grid)  simulations produced using the cosmological code {\enzo} \citep[][]{enzo14}. {\enzo} is a highly parallel code for cosmological (magneto)hydro-dynamics, which uses a particle-mesh N-body method (PM) to follow the dynamics of the Dark Matter (DM) and a variety of magneto-hydrodynamical (MHD) solvers to evolve the gas component on a support uniform or adaptive grid \citep[][]{enzo14}.  
 
\subsection{How complex is the formation of a galaxy cluster?}
\label{amr}

Sitting at the top of the mass distribution of cosmic structure, galaxy clusters are a key astrophysical object which  form and evolve over cosmological timescales ($\sim 1-10 \rm ~Gyr$). They behave under many respects as "closed boxes", whose properties are strongly linked to cosmology\citep[e.g.][]{2012ARA&A..50..353K}. Given their large volume and overdensity, they usually represent the first detectable signpost of the cosmic web at most wavelengths \citep[e.g.][]{2002ARA&A..40..539R,2019SSRv..215...16V} as a well as the perfect plasma laboratories in the Universe \citep[][]{bj14}.

The simulation presented in this work includes the effect of magnetic fields, radiative cooling of gas and energy feedback from active galactic nuclei in a standard $\Lambda$CDM cosmological setup. Adaptive mesh refinement (AMR) was used to selectively increase the dynamical resolution in the formation region of galaxy clusters, up to $\Delta x_{\rm max}=32 \rm ~kpc$ (comoving),  which is mandatory to resolve turbulence and magnetic field amplification. \\

I re-simulated the objects studied in here in two flavors: a) with non-radiative runs, only including gravity, hydrodynamics and magnetic fields or b) with radiative runs, including equilibrium gas cooling (assuming a primordial chemical composition) and thermal/magnetic feedback from active galactic nuclei (see Sec.\ref{agn}). The magnetic field has been initialised to $B_0=10^{-10} \rm G$ everywhere in the box at the start of the simulation. More details can be found in \citet[][]{va17info}, \citet{wi17} and \citet{2018Galax...6..128L}.

All physical fields of the simulations were outputted on disk every $\Delta t \sim 3.11 \cdot 10^{6}$ yr and at the maximum available resolution ($\Delta x_{\rm max}=32 \rm ~kpc$).\\ 
 
\subsubsection{The morphology of complexity}
\label{cluster}
\begin{figure}
\includegraphics[width=1.2\textwidth]{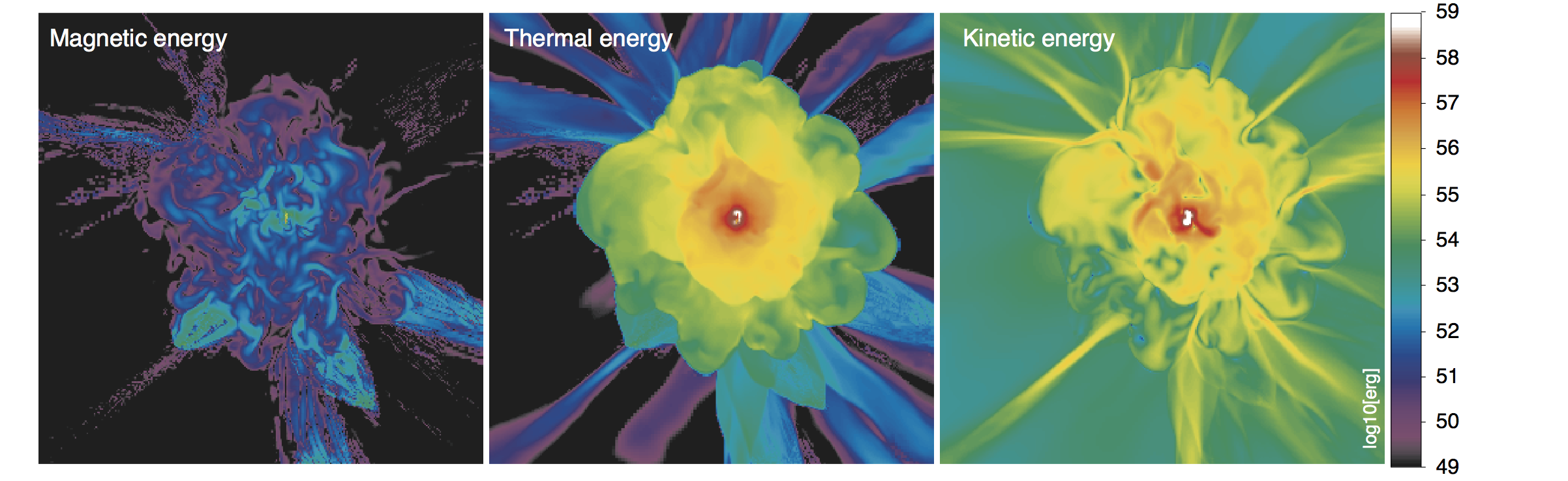}
\caption{Magnetic, thermal and kinetic energy  for a slice through the centre of a $\sim 3.0 \cdot 10^{14} M_{\odot}$ cluster at $z=0.01$. The panel is $6 \times 6 \rm Mpc^2$ across. }
\label{fig:map0}
\end{figure}

Fig.~\ref{fig:map0} shows the spatial distribution of thermal, magnetic and kinetic energy for a thin slice crossing the centre of one simulated galaxy cluster (with a final virial mass $M_{\rm 100}\sim 3 \cdot 10^{14}M_{\odot}$)  at $z=0.01$.
 
The thermal and the kinetic energy fields (and to a smaller extent, also the magnetic energy) closely follow the roughly spherically symmetric distribution of gas density, which reaches $\approx 5 \cdot 10^{-3} \rm ~part/cm^3$ in the cluster core and $\sim 10^{-5} ~\rm part/cm^3$ in the cluster periphery, which are typical values for galaxy clusters. 

The thin slice shown here allows to see smaller scale perturbations associated to the various mechanisms responsible for the mass growth of the cluster (which is still ongoing at $z=0$). 
  
Several filaments are connected to the cluster periphery, as well as sharp boundaries, and they penetrate the quasi-spherical envelope of  strong shocks  (i.e. Mach number $M \geq 10-100$)  at which the infall gas  kinetic energy gets thermalized \citep[e.g.][]{ry03,pf06}. In such peripheral regions, the flow is predominantly supersonic, with accretion velocities exceeding the local sound speed. 
 
 Within the denser cluster atmosphere, the budget of kinetic energy gets smaller than the thermal energy, 
 and velocity fluctuations are due to residual subsonic motions, mostly of turbulent origin \citep[e.g.][]{do05,in08,va11turbo}.  The magnetic energy is subdominant everywhere in the cluster, and barely reaches a few percent of the thermal/kinetic energy within the cluster, remaining of order $\leq 10^{-4}$ of the thermal energy in most of the volume\footnote{It shall be remarked that the magnetic field is by far the energy field which is most sensitive to changes in the spatial resolution, because of the strong dependence of the dynamo amplification with the numerical Reynolds number \citep[see discussion in][]{review_dynamo}. Due to the limited spatial resolution probed in this simulation, the simulated magnetic field is a factor $\sim 10$ smaller than suggested by radio observations \citep[e.g.][]{bo13}.}

\begin{figure}
\includegraphics[width=1.2\textwidth]{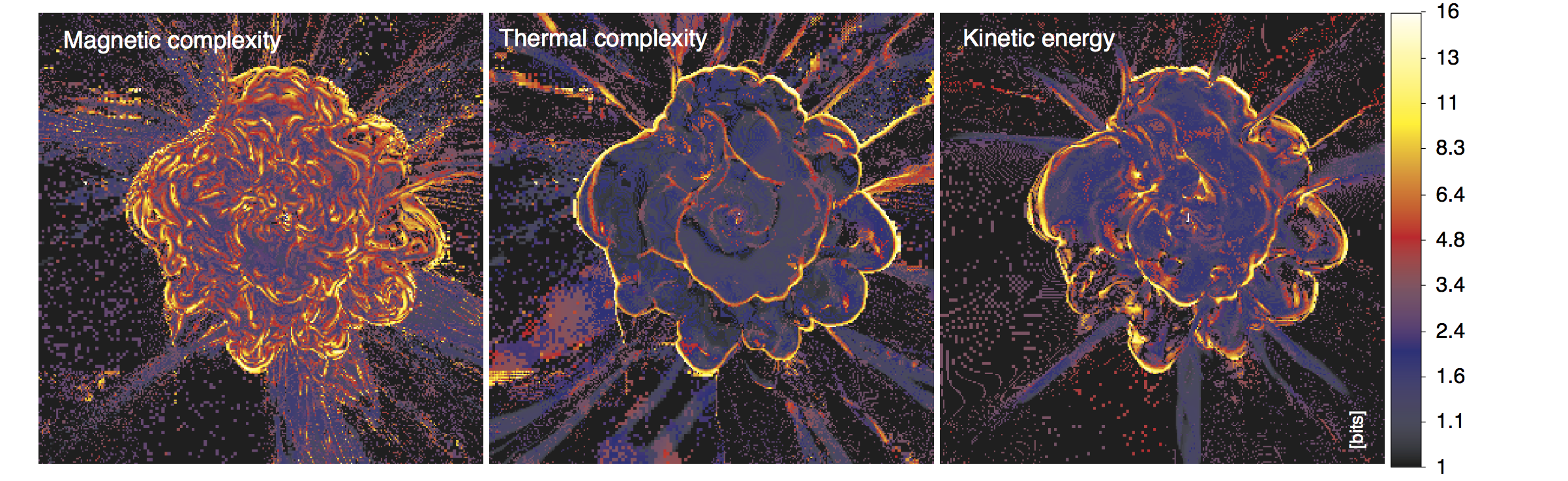}
\caption{Magnetic, thermal and kinetic complexity for the same cluster and selection of Fig.\ref{fig:map0}.}
 \label{fig:map1}
 \end{figure}

The spatial distribution of  statistical complexity for the same cluster is given in Fig.~\ref{fig:map1}. To compute $C_\mu$  I employed  $N_{\rm bin}=200$ logarithmic energy bins,  and compared the outputs of two snapshots separated by one root grid time step ($\Delta t \sim 3.11 \cdot 10^{6} ~\rm yr$). 

The complexity distribution captured by the statistical complexity filtering shows a wide distribution of scales that highlights both large scale complexity pattern associated to several filamentary accretions in the cluster periphery, as well as small-scale fluctuations in the innermost cluster regions. Patches of significant complexity are manifestly associated with large physical jumps of the energy fields,  mostly concentrated  in narrow zones near shocks (as it can be independently measured with shock finder methods). 
The thermal energy, on average, requires $\sim 5-10$ times more information (reflected in a higher $C_\mu$ ) because the jump of thermal energy in strong shocks is much larger than that of kinetic energy. 
Although non-radiative numerical shocks obey  ``simple'' Rankine-Hugoniot jump conditions, at every time step only a small fraction of the cells within a specific  energy bin (i.e. environment)  is modified by shocks. To predict whether or not a specific cell is going to be affected by shocks, or more in general how its energy should evolve, additional information is required to solve the local Riemann problem (on average, of the order of $\sim 10 ~\rm bits/cell$ here) . 

 Compared to the thermal energy, the kinetic energy $E_K$ shows more complex fluctuations away from shocks, as well as closer to the cluster centre. The ICM is known to host volume filling subsonic turbulence at all epochs, as result of gravity-driven random motions \citep[e.g.][]{va11turbo}; thus most of such fluctuations are due to  turbulence on the short timescale which separates the two time steps used for this analysis. 
Finally, the magnetic energy is sub-dominant compared to the thermal/kinetic energy of the ICM. However, this means that one needs more information  in order to predict the evolution of magnetic fluctuations.  The ultimate driver of its evolution is indeed turbulence, through  small-scale dynamo amplification \citep[e.g.][]{xu09,2019MNRAS.486..623D}: this makes the evolution of the magnetic typically more complex at all scales and distances from the cluster centre.  
Very similar results were found for all simulated galaxy clusters using {\enzo}, provided that small differences in the dynamical state and in the shock history of each object are reflected in the final distribution of complexity \citep[e.g.][]{va17info,va19info}.

\begin{warning}{Why is complexity useful to simulate the formation of galaxy clusters}

If complexity can be measured at run-time while the simulation is running, 
this information can allow the simulation code 
 to identify exactly where and when complex evolutionary patterns are emerging
 in the domain. Coupled with adaptive mesh refinement, this approach can selectively increase the local force and spatial resolution at run-time.
Traditionally, this task is performed by fixing a-priori some relevant threshold values for the combination of several quantity of interest (e.g. matter overdensity, local jumps in thermodynamical quantities, etc), and letting the simulation reduce (typically, by halving) the local mesh resolution whenever such threshold values are exceeded.
Examples of this include refining on the local matter overdensity  \citep[e.g.][]{1998ApJ...495...80B,2010MNRAS.401..791S}, on the velocity shear \citep[][]{2006ApJ...638L..25K}, on gas vorticity \citep[][]{in08}, on 1-dimensional velocity jumps tracking shocks \citep{va09turbo}, on magnetic field intensity \citep[][]{xu09}, etc. Each of these choices depends on the simulator's prior knowledge and expectations about the physics in the simulation, and it is thus biased to refine on behaviours that can be predicted (or at least guessed) before the run begins. Moreover, dedicated tests have shown that the employment of too aggressive AMR techniques introduces un-physical perturbations to simulated systems \citep[e.g.][]{2015A&C.....9...49S}. 

The possible application of statistical complexity as a method to trigger new refinements during the simulation can offer an unbiased way of improving calculations, independently on the observer's prior expectations on the problem under study.

Moreover, the usual workflow requires first to run low or moderate resolution versions of such simulations, and to apply increasingly aggressive AMR in a second step, by restarting the first run from an interesting epoch. 
Information Theory thus allows a more elegant solution to this challenge (which can be a challenge for large simulations),  
because statistical complexity only relies on the symbolic analysis of the data-stream generated by the simulation, without any a-priori knowledge of what is in the data, nor of the relevant physical threshold to exceed. 

\begin{figure}[b]
\sidecaption
\includegraphics[width=0.49\textwidth]{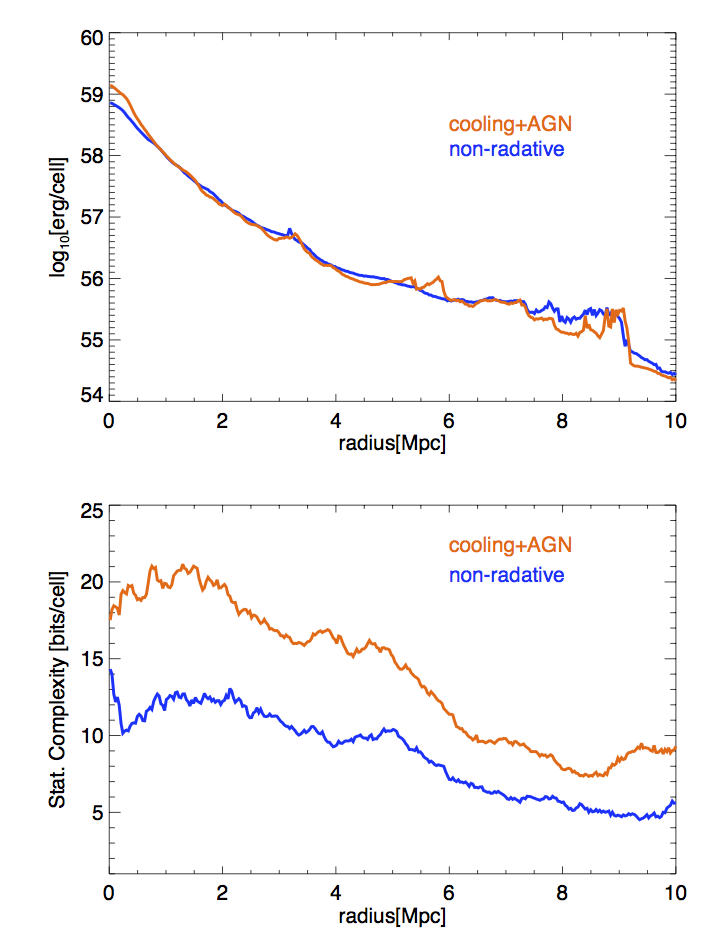}
\caption{Top panel: radial profile of the average total (thermal, kinetic and magnetic) energy for  a simulated galaxy cluster at $z=0.01$, in a non-radiative setup or in a model including radiative cooling and AGN feedback. Bottom panel: radial profile of the average statistical complexity of all energy fields for the same two resimulations.}
 \label{fig:prof}
\end{figure}
\end{warning}

\subsubsection{The impact of galaxy formation on cluster complexity}
\label{agn}

The impact of galaxy formation  physics on the dynamical evolution of the intracluster medium (namely radiative cooling leading to the collapse of overdense gas clumps and their later feedback on the surrounding gas distribution via feedback) is a wide field of research for numerical simulations, as it significantly modifies the (self-similar) scaling relations between the total mass of clusters and their thermodynamical or observable properties \citep[e.g.][]{2007ApJ...668....1N,teyssier11}. 

I investigated the role of galaxy formation on the emergence of complexity with variations of the physical model discussed in the previous Section. 
In detail, radiative re-simulations of the same galaxy clusters include equilibrium gas cooling from primordial chemical composition, and effective model for the (large-scale) energy release by AGN events, in the course of the simulation. In this case, each feedback event (triggered whenever the gas density within a cell exceeded $10^{-2} \rm part/cm^3$) releases $10^{60}  \rm erg$ of thermal energy and $10^{59} \rm  erg$ of magnetic energy (as bipolar structure). The above simplistic model  bypasses the problem of following prohibitively small scales involved in the accretion of gas onto super massive black holes. However, it has been shown to adequately reproduce the thermodynamical properties of the observed ICM on $\geq 100 ~\rm kpc$ scales \citep[e.g.][]{va13feedback}. 

The combined effect of cooling and AGN feedback on simulated clusters is typically to increase the gas density in the core, to produce large transients in gas temperature, to promote the significant expulsion of baryons from the innermost cluster regions, as well as to introduce more turbulence and clumpiness in the ICM \citep[see discussion in ][]{va13feedback}.

As an example, Figure \ref{fig:prof} shows the average radial distribution of the total energy (kinetic, thermal and magnetic) for the gas in the simulated cluster of Fig.\ref{fig:map0} at $z=0.01$. On the one hand, the overall radial energy distribution is similar to the non-radiative re-simulation, but the radial profile shows more substructures (due to the enhanced clumping of gas) as well as a higher energy budget in the cluster core (as a mixed effect of gas compression and extra heating from the central AGN). A more detailed analysis of re-simulations with AGN feedback can also be found in \citet{va17info}. 
Overall, despite the additional physics included in the simulation, the volumetric distribution of all fields is quite similar for $\geq $ $\rm Mpc$ radii, indicating that the equilibration of non-gravitational perturbations within the cluster atmosphere is overall efficient enough to erase most of the signatures from AGN across the entire cluster volume.  

Remarkably, the radial distribution of total statistical complexity (bottom panel of Fig. \ref{fig:prof}) shows large differences across the entire radial extent, out to $\sim 10$ $\rm Mpc$, leading to an overall $\sim 50-100 \%$ increase of $C_{\rm \mu}$ everywhere. All energy fields show an almost equally increased complexity, with a larger difference in all cases for the innermost cluster regions. 
The fact that more complexity is found even at large radii suggests that the extra complexity is not entirely due to the central AGN in the cluster, but that it  probably was contributed by the activity of several AGN in the volume, and/or by volume filling processes produced before the cluster was fully assembled.

\begin{figure}[b]
\sidecaption
\includegraphics[width=0.6\textwidth]{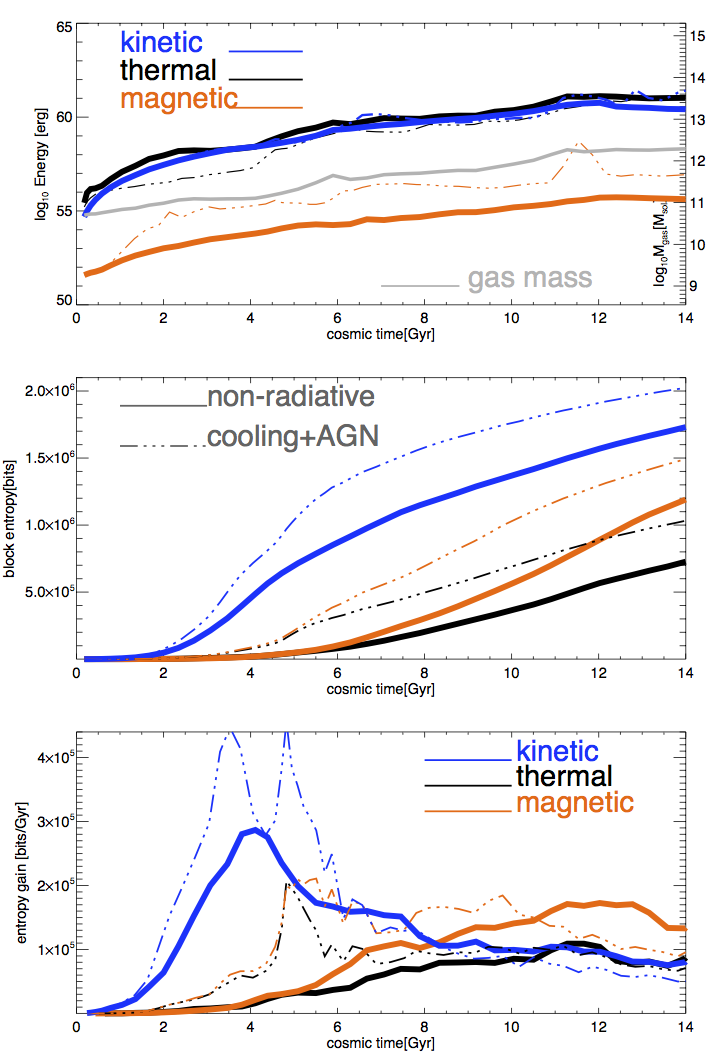}
\caption{Top panel: Evolution of the kinetic, thermal and magnetic energy fields (as well as of the gas mass) for a sample of $1.9 \cdot 10^5$ cells in the central region of the galaxy cluster of Fig.1-2. The thick lines refer to the non-radiative simulation while the thin lines are for the radiative simulation including feedback from AGN. Central panel: evolution of block entropy for the same selection of cells.  Bottom panel: evolution of the entropy gain for the same selection of cells.}
 \label{fig:block_entropy}
\end{figure}
\begin{warning}{Why is complexity useful to study the evolution of intergalactic gas}

When and how did the extra complexity in the cluster arise? 

The block entropy analysis introduced in Sec.~\ref{subsubsec:block_entropy}, is a monotonically increasing parametrisation of the information content of a system, hence it is very suitable for time-integrated studies, in which past and recent events can be dynamically related. 

Following \citet{va17info}, I computed the  block entropy, $H(L)$, and its source rate term, $h_\mu(L)$, for the entire sequence of kinetic, thermal and magnetic energy from $z=30$ to  $z=0$ within a volume centre on the formation region of a cluster. The full analysis of the sequence of symbols (e.g. energy levels) computed by the simulation requires a huge amount of data (e.g. $\geq 2$ Tb to follow all cells in the simulation at high resolution for the entire sequence of $440$ root grid timesteps), hence the analysis is here restricted to a small representative data set in cluster formation region, comprising $\sim 1.6 \cdot 10^5$ cells.  More optimised algorithms thus need to be developed for the full analysis of the entire data flow of existing and future large simulations.
While the absolute value of block entropy at a specific epoch may depend on the volume being investigated (as well on specific choices of the binning of energy levels and on the time sampling frequencies) the relative growth of  block entropy in the energy fields is more robust to model variations \citep[e.g. see][for a discussion]{va17info}.\\

Figure \ref{fig:block_entropy} (top panel) gives the evolution of the total energies contained in the 
selected sub-region in the cluster, from $z=30$ to $z=0$.  By the end of the simulation, the total thermal and kinetic energy of gas are nearly identical, as implied by the radial profile while in the first half of the simulation (and before the efficient heating by AGN) the thermal energy in the radiative simulation is lower, due to the effect of radiative losses on the densest clumps in the region.  On the other hand, the magnetic energy in the same region is larger in the radiative run, due to the due to the combined effect of gas compression (induced by cooling) and of the additional  magnetisation induced by AGN feedback.

The middle panel of Figure \ref{fig:block_entropy} gives the evolution of block entropy for the kinetic, thermal and magnetic energy of both re-simulations as a function of time. In this case, larger differences are visible:  
the block entropy increases in a monotonic way, as expected,
 reaching $\langle H(L) \rangle \approx 22.5$ bits/cell in the non-radiative case  and   $\langle H(L) \rangle \approx 28.1$  bits/cell in the radiative run. Despite the similar thermal and kinetic energy distribution across most of the simulation, the complexity of all energy fields is consistently larger in the radiative run. 
 Cooling and feedback have overall a little impact on the relative complexity of the energy fields after the cluster assembled, for $t \geq 4 ~\rm Gyr$, and their role is more evident at earlier times.
 Longer before contributing to the mass of the  cluster in this region at late epochs, the cosmic gas in the simulation was subject to a complex dynamical evolution of all fields, long before there was (approximate) equilibrium between the forming gravitational well of the cluster and the thermal gas energy.

The entropy gain, $h(L)$ (bottom panel of Fig.\ref{fig:block_entropy}) better illustrates when and how complexity gets increased in the two runs: shortly after mergers and matter accretions experienced by the cluster (e.g. see the spike in gas mass at $t \sim 6 \rm ~Gyr$), as well as after AGN bursts inducing outflows  when this is included  (e.g. bursts at $t \sim 3\rm ~Gyr$, $t \sim 5 \rm ~Gyr$). Since the local fluctuations driven by AGN in all fields are more violent than in mergers, the entropy gain also is manifestly more significant after AGN bursts.  As a consequence, the largest spikes in entropy gain are reached well before the cluster has fully assembled, i.e. for $t \leq 5 ~\rm Gyr$, when its gas mass was $\leq 10\%$ of its $z=0$ value and a large fraction of infall kinetic energy still had to be virialized. 

This test well illustrates the power of Information Theory applied to astrophysical simulations, in which several different mechanisms operate together: complexity analysis can detect and expose large differences related to the underlying complexity of the adopted physical models being tested, even when 
detecting such differences is difficult with standard analysis. For example, the impact of AGN on the kinetic perturbations away from cluster cores is hard to detect in simulations \citep[e.g.][]{ka07,dubois11,va13feedback,2017MNRAS.467.3827P}, owing to the rather fast dissipation of turbulent motions in the ICM. \\

\end{warning}

\begin{figure}[b]
\sidecaption
\includegraphics[width=0.49\textwidth]{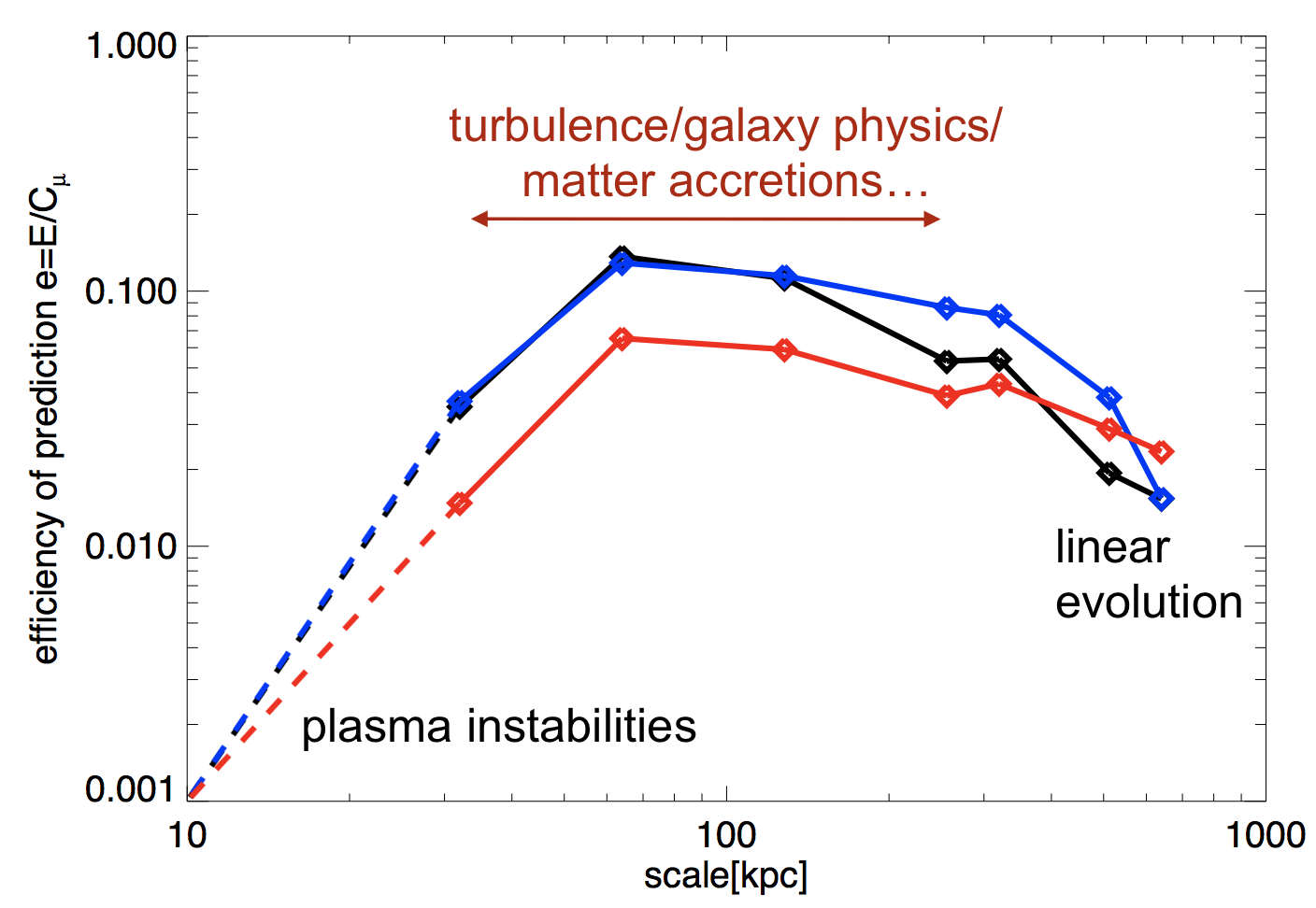}
\caption{Relation between the efficiency of prediction (Sec.\ref{subsubsec:eff}) at different interpolation scales, for sample of $1.9 \cdot 10^5$ cells in a forming galaxy cluster. The dashed lines connects the values of $e$ measured in the simulation, with the $e$ values estimated from plasma physics on unresolved scales (see text for explanation). The black line is referred to the thermal energy, the blue and the red to the kinetic and magnetic energy, respectively.}
 \label{fig:efficiency}
\end{figure}

\begin{warning}{Which scales contains most information?} 

\label{eff_icm}

The volume of galaxy clusters is large, and it comprises so many different spatial scales that it is natural to ask whether there is there a preferred scale at which the emergence of complexity is maximum, and which is the scale that contains the most information on the evolution of such systems. 

In \citet{va17info} I studied the evolution of energy fields in $\approx 1.9 \cdot 10^5$ cells in the centre of a forming galaxy cluster, computing their average $H(L)$ and $E(L)$, for different levels of linear coarse-interpolation of the data, from the coarsest $\Delta x_{max}=634$ kpc resolution to finest  $\Delta x_{min}=32$ kpc one. 
Figure \ref{fig:efficiency} gives the trend of efficiency, $e$, measured in the galaxy cluster simulation as a function of the adopted interpolation scale, which displays a similar
trend for all energy fields. 

The maximum $e$ is found in the range of $\sim 63-190$ kpc, with 
$e \approx 0.1-0.2$. 
This range of scale is the typical one of turbulent eddies in the simulated ICM \citep[e.g.][]{va12filter,sch16}, of the typical outer correlation scale of observed and of simulated magnetic fields in the ICM \citep[e.g.][]{xu09,bo13}, as well as of  measured projected density fluctuations in X-ray \citep[e.g.][]{ga13}. It appears therefore reasonable that $e$ is the largest on  where the ICM presents the highest degree of dynamical self-organisation,  which is also routinely targeted by telescope observations at different wavelengths. On the other hand, the coarse-grained evolution on much larger scales allows  a more robust prediction of future evolution as on such scales the evolution approaches the linear regime of small density perturbations. However, this also makes such evolution "easier" to predict, making the $E/C_{\rm \mu}$ ratio lower than for smaller scales. \\

This simulation cannot probe scales $\ll \Delta x_{min}=32$ kpc, however a few basic considerations suggest that $e$ should decrease again for such "microscopic" scales. 
The 
efficiency of prediction at these scales can be estimated by considering that the dynamics of thermal gas in the ICM can be assumed to be a Markovian process, which means that the thermodynamical value of single particles only depends on their last microstate. Hence $E=C_\mu-L h_\mu \approx C_\mu-h_\mu$ because $L \approx 1$ \citep[][]{2004PhRvL..93n9902S,prokopenko2009information}.

In this regime, the thermodynamic entropy also gives the statistical complexity, which for the thermal particles of the ICM is $S \sim 10 \rm keV/particle$ \citep[e.g.][]{borgani08}. On the other hand, the source entropy rate crucially depends on how energy is exchanged between particles on very small scales. The ICM plasma is expected to be weakly collisional on these scales, hence energy gets mostly exchanged by collective plasma effects (including a wide range of possible plasma instabilities) acting on $\sim $ seconds timescales  \citep[e.g.][]{2011MNRAS.410.2446K,bl11}. In this scenario, the extremely fast action of plasma collective implies that on microscopic scales the efficiency of prediction is $\approx 1$ only in the scale of {\it seconds}, while it rapidly drops to zero for any other longer timescale.
Even in the rather standard (and probably out-dated) model of a collisional ICM \citep[][]{SA88.1}, where  Coulomb collisions between particles solely exchange energy, the entropy rate can be estimated to $h \approx 10^{-7} \rm keV/particle/yr$ based on the expected proton-proton Coulomb collision frequency, implying that $e \approx 1$ only for $\leq 10^{5}$ $\rm yr$ timescales. 

In summary,  a detailed thermodynamical view of single particle interactions in the ICM appears to irrelevant to predict the evolution of the ICM on any astronomically relevant scale, given the enormous difference in scale between microscopic and macroscopic processes involved. Collective processes emerge on $\geq $ $\rm kpc$ scales, which are routinely observed by telescopes and usually are simulated with numerical simulations represent the best range of scales at which the "emergent" properties of the ICM are evident, and where the evolution of such systems can be effectively described using a (magneto) hydrodynamical model.
In particular, the $\sim 50-200$ $\rm kpc$ range of scales appears to be the one that maximises the efficiency of prediction in the investigated cosmological simulations, and that is used in the subsequent investigation of complexity in the entire cosmic web.

\end{warning}

\subsection{How complex is the formation of the cosmic web?}

With a second set of numerical simulations, I measured the distribution of complexity in a full cosmological volume $(40 ~\rm  Mpc/h)^3 \approx 57^3 \rm ~Mpc$, simulated with {\enzo} using $400^3$ cells and DM particles, at the constant resolution of $\Delta x=141$ kpc/cell.  With this setup,  I investigated several variations of gas physics and of cosmological parameters in order to assess their impact on the emergence of complexity, as detailed in \citet{va19info}. 
 All simulations employed the numerical MHD scheme of Dedner \citep[][]{ded02} as in the previous case, with a magnetic field initialised to be  $B_0=0.1 ~\rm nG$ (comoving) along all magnetic field components at the begin of the simulation ($z=40$). \\
 
The choice of this spatial resolution was motivated by the measured trend of the efficiency of prediction  Sec.\ref{eff_icm}, which 
ensures the most "information rich" view of the emerging complexity of the cosmic web. 

The statistical complexity, $C_{\rm \mu}$, is here measured as in \citet{vazza19} (and similar to the previous Section), by employing 
 equal logarithmic energy bins , ranging from the maximum and the minimum of each energy field, respectively, and considering a time spacing of $dt=5$ timesteps ($\approx  200 ~\rm Myr$) between snapshots. 

\begin{figure*}
\sidecaption
\includegraphics[width=0.97\textwidth]{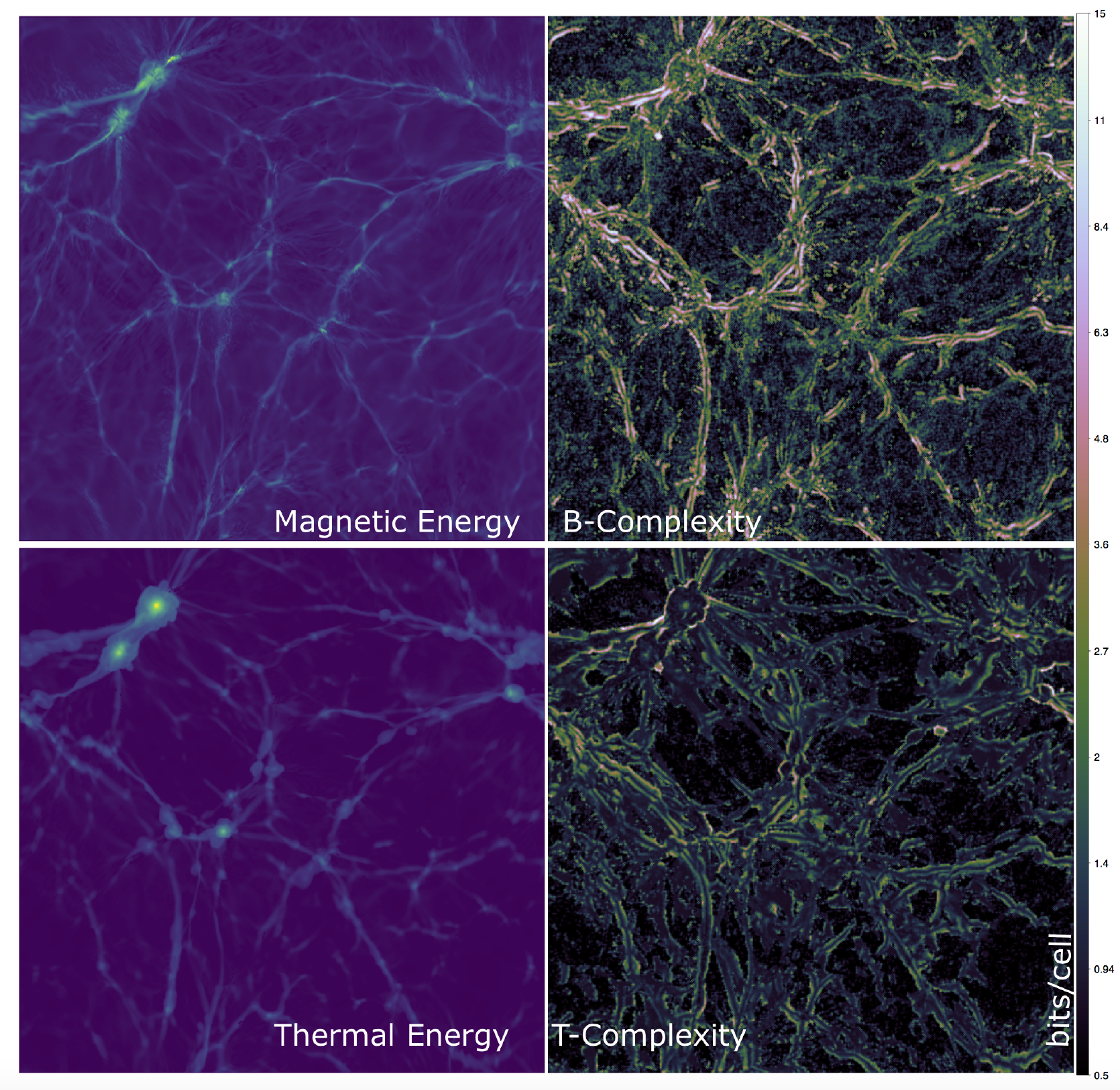}
\caption{Slices through a $57^3 \rm ~Mpc^3$ simulated volume at $z=0$. The top panels show the magnetic energy (left) or the magnetic complexity (right), the lower panels show the thermal energy (left) or the thermal complexity (right).}
 \label{fig:large0}
\end{figure*}

\begin{figure*}
\sidecaption
\includegraphics[width=0.45\textwidth]{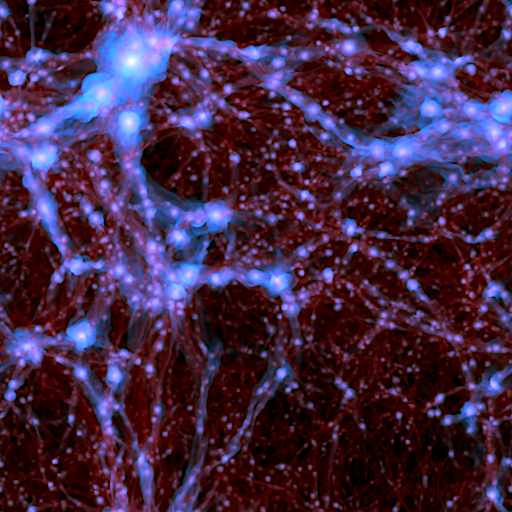}
\includegraphics[width=0.45\textwidth]{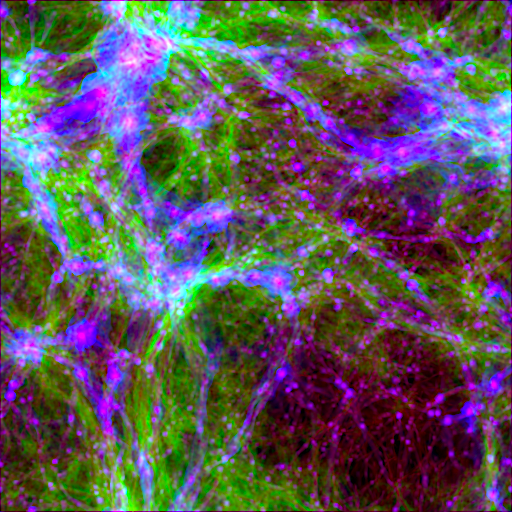}
\caption{Left panel: total gas density (red) and gas temperature (blue) for the $57^3 \rm ~Mpc^3$ simulated volume at $z=0$. Right panel: total complexity (green) over imposed to the previous map.}
 \label{fig:rgb}
\end{figure*}

A couple of visualisation examples of $C_{\rm \mu}$ for the full cosmological volume, and of its spatial relation with the entire cosmic web on scales much larger than the ones probed in the previous cluster analysis (Sec.\ref{amr}) are given in Fig. \ref{fig:large0} and \ref{fig:rgb}. 

Figures \ref{fig:large0} give a 2-dimensional view of the distribution of the thermal and magnetic energy fields for a thin slice through the simulation at $z=0$, and the corresponding distribution of statistical complexity for the same volume. 
A 3-dimensional rendering of the entire simulated volume at the same epoch is given in 
Fig.\ref{fig:rgb}, which shows the total gas density in red,  the gas temperature  in blue, and additionally the total complexity (in green) in the right panel.

While the large-scale distribution of all energy fields closely  trace the matter distribution of the cosmic web and its related gravitational potential, with maxima located in self-gravitating matter halos, the spatial distribution of complexity appears broader. This means that, across the full range of cosmic environment, regions of significant different energy may have an equally complex evolution, depending on their local dynamics and past history. The visual inspection also shows that prominent spikes of complexity are associated with shocks, marked as sharp contours around filaments or at the periphery of halos in the volume, in line  with the previous Section.

\begin{figure*}
\includegraphics[width=0.33\textwidth]{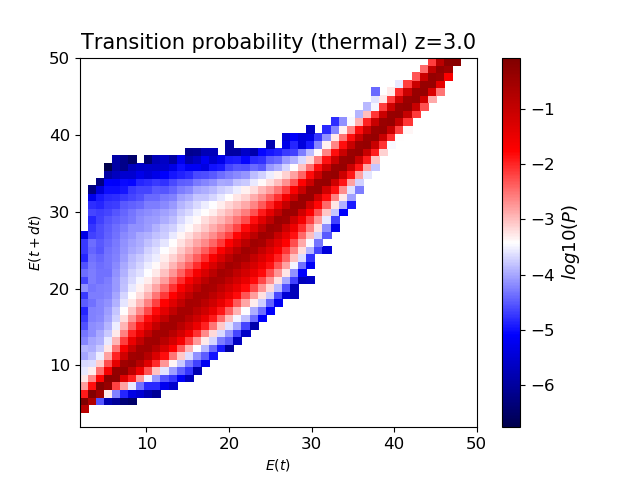}
\includegraphics[width=0.33\textwidth]{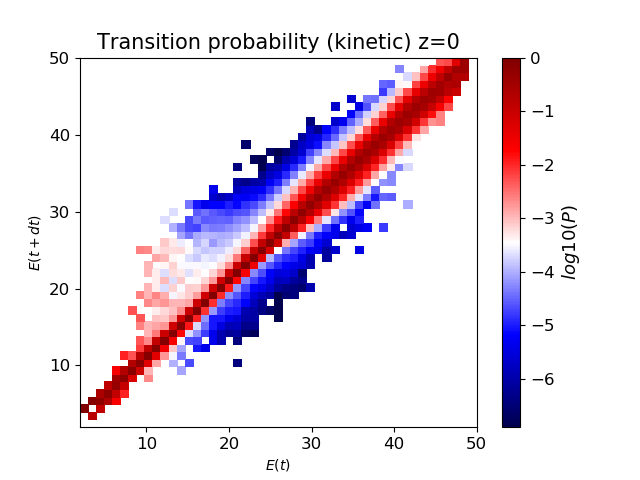}
\includegraphics[width=0.33\textwidth]{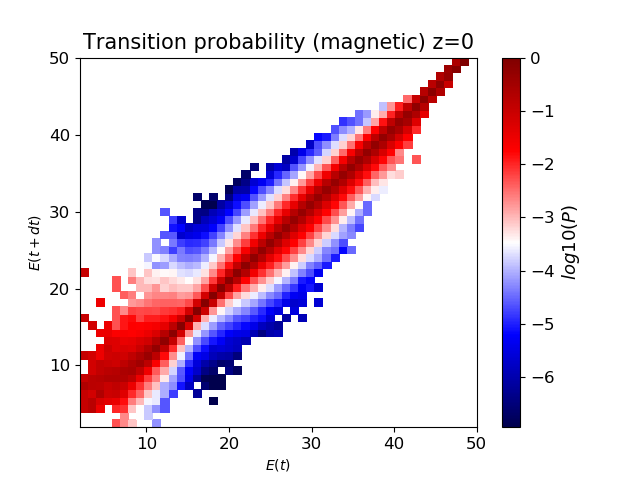}
\caption{Matrix of transition probabilities, $P_{xyz}$ measured between the energy states at timestpes $t$ and $t+dt$, considering transitions of thermal (left), kinetic (centre) and magnetic (right) energy at $z=0.0$.}
\label{fig:prob}
\end{figure*}

The complexity in different environments directly follows from the transition probability matrix (Sec.\ref{stat_complex}) across the entire range of cosmic overdensities. For example, the 3-dimensional structure of the complexity  traces shocks around filaments and massive halos, for which a large spread in $P(E_2|E_1)$ can be expected.
Fig.\ref{fig:prob} gives the transition probability matrix, $P(E_2|E_1)$ (see Sec.\ref{stat_complex}) measured in the full volume at $z=0.0$ and referred to 50 logarithmical energy bins. 
Here the diagonal 1-to-1 relation corresponds to little complexity
transitions, in which $E_i(t)$ states are mapped onto the same 
level at the following timestep.  On the other hand, a large 
spread around the 1-to-1 correlation hints at complex transitions which require more information to predict, like large thermodynamic jumps associated with shocks (see also Sec.\ref{cluster}).

Indeed, most energy levels in the intermediate range are spread in the probability distribution. This is consistent with the fact that strong structure formation shocks typically change the energy content of gas particles in the linear overdensity regime, with $T \sim 10^4 \rm ~K$ (mostly related to the most filamentary part of the cosmic web), even on the short timescale of the simulation timestep. On the other hand, as observer also in Sec.\ref{cluster}, 
strong shocks are able to cause only smaller transitions of kinetic and magnetic energy levels within cells, following from shock jump conditions.
Across most environments, we can observe a spread of magnetic energy levels in the probability matrix, mostly associated with the fact that the magnetic field can be changed both by compression or by magnetic field amplification, via small-scale dynamo -  albeit at a rate limited by the modest spatial dynamical range that is achieved by turbulence and dynamo in this simulation \citep[e.g.][]{review_dynamo}.

\begin{figure*}
\includegraphics[width=0.245\textwidth]{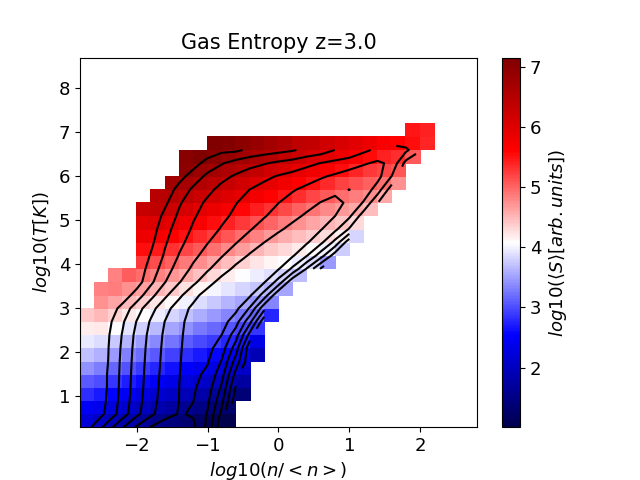}
\includegraphics[width=0.245\textwidth]{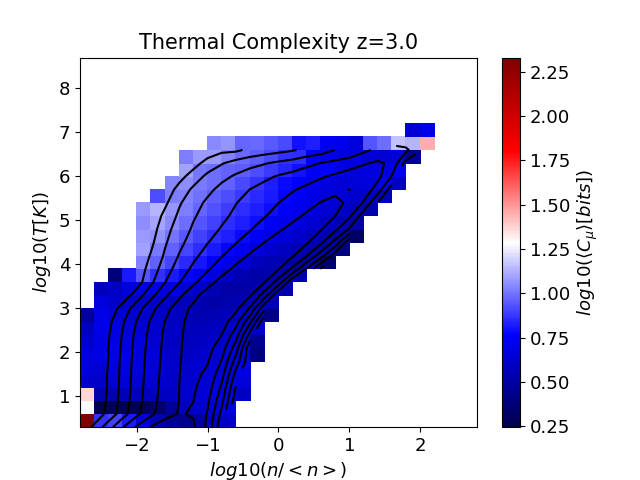}
\includegraphics[width=0.245\textwidth]{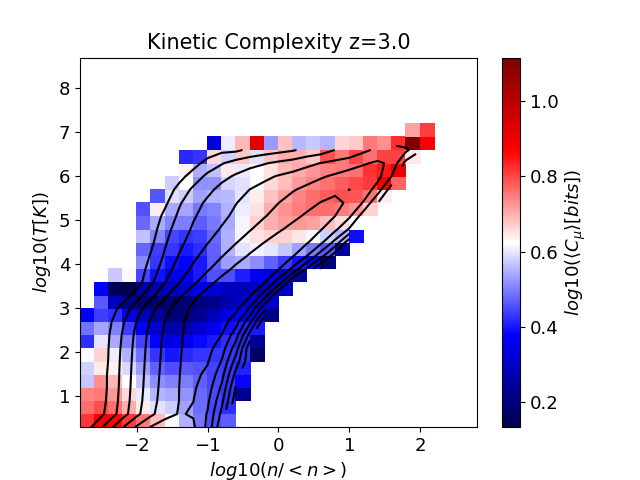}
\includegraphics[width=0.245\textwidth]{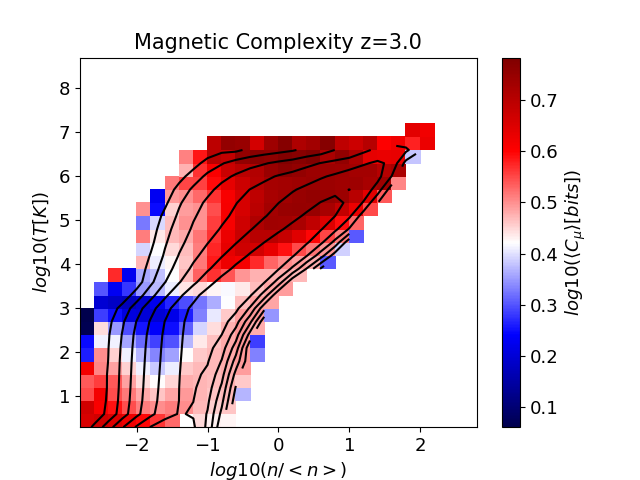}
\includegraphics[width=0.245\textwidth]{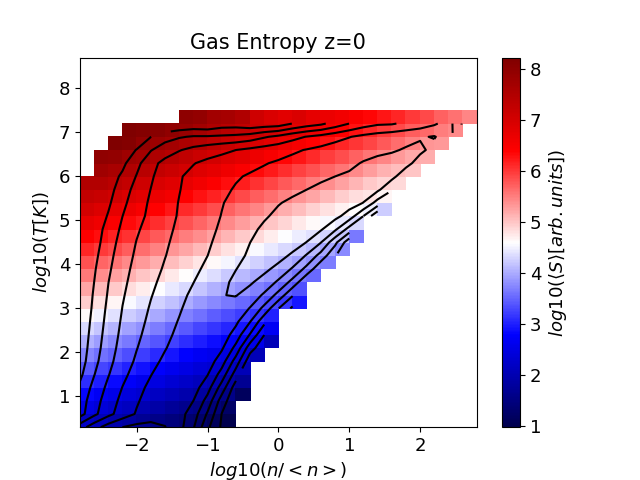}
\includegraphics[width=0.245\textwidth]{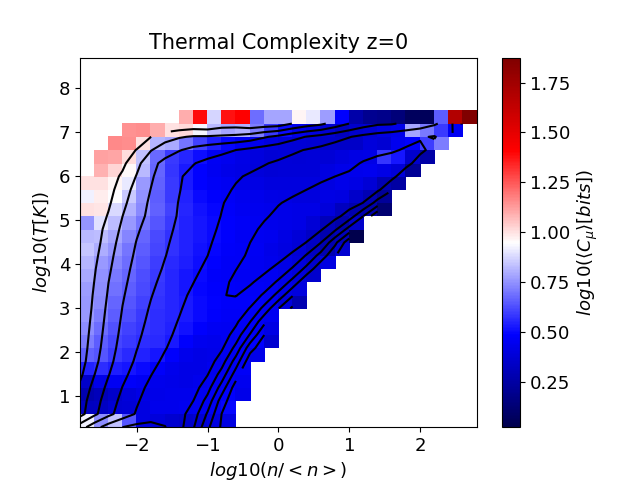}
\includegraphics[width=0.245\textwidth]{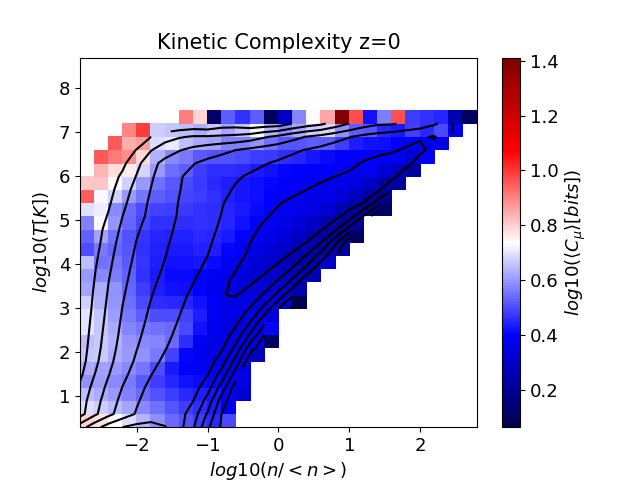}
\includegraphics[width=0.245\textwidth]{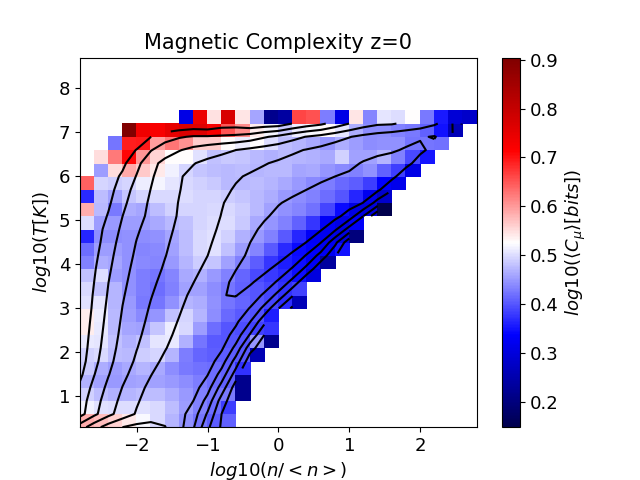}
\caption{Phase diagrams showing the average complexity of the thermal, kinetic and magnetic energy at $z=3$ and at $z=0$ for the simulated cosmic web. The first column shows the average gas entropy for the same boxes. The grey contours in each panel show the gas density distribution (with logarithmic spacing of contours).}
\label{fig:phase1}
\end{figure*}

The distribution of complexity as a function of the cosmic environment is better quantified by the phase diagrams in Fig.\ref{fig:phase1}, which give the average statistical complexity of energy fields, for the reference epochs of $z=3$ and $z=0$ as a function of $\rho$ and $T$. Approximately, the range of $n/\langle n \rangle \geq 10-100$ marks here halos of groups or clusters of galaxies while $T \leq 10^4 \rm~ K$ marks cosmic voids.  Intermediate ranges of values are the location of linear or mildly non-linear structures of the cosmic web, i.e. matter sheets and filaments. 

From the phase diagrams, we observe that the peak of complexity in the cosmic volume moves across the environment as a function of time. 

At $z=3$, most of halos in the simulation are still being assembled, leading to a significant conversion of infall kinetic energy into thermalisation and magnetic field amplification, prior to the establishment of (approximate) hydrostatic equilibrium. The complexity thus peaks for the high densities at which the conversion of infall kinetic energy is more prominent, i.e. in halos undergoing mass growth, typically reaching $C_{\rm \mu} \sim 10^2$ bits/cell.

Later on,  it is the periphery of galaxy clusters or filaments to become the most complex environment in the volume, with  $\langle C_{\rm \mu} \rangle \geq 10-10^2$ bites/cell. Conversely,  the internal volume of halos becomes less complex ($\langle C_{\rm \mu} \rangle \sim  10$ bits/cell) by $z=0$,  because only rare single perturbations (e.g. major mergers) can change pre-existing energy levels by a large amount. Despite the significant change in the spatial resolution, these trends are in line with the highest resolution of galaxy clusters view previously discussed in  Sec.\ref{amr}.

Finally, the average complexity is  small in voids, $\langle C_{\rm \mu} \rangle \ll 10$ bits/cell,  due to their relatively simple evolution, mostly ruled by adiabatic gas expansion. Nevertheless, a residual amount of complexity is also found in low density regions,  resulting from the expansion of the structure formation shocks released during the very first stage of halo formation, and still expanding into lower densities.\\

The first column of Fig.\ref{fig:phase1} additionally shows the phase diagram of gas entropy ($S \propto T/\rho^{2/3}$), which overall is similar to the one of thermal complexity. The latter is not surprising considering that there is indeed a duality between entropy and Shannon information (as defined in Eq.\ref{shannon}), which stems from the basic definition of entropy in statistical thermodynamics:

\begin{equation}
    S=-k_B \sum P_i \log P_i
    \end{equation}
  (where $P_i \propto e^{\frac{\epsilon_i}{k_B T}}$ is the probability of the energy state $\epsilon_i$ with temperature $T$, in a Boltzmann distribution).

In the non-radiative simulations considered here, the gas entropy in the volume is only increased by the irreversible dissipation at shocks, or by spurious numerical effects. The fact that the high-temperature envelope of the phase diagrams corresponds to the location of maximum entropy production and maximum complexity confirms that dissipative processes related to structure formation are indeed the main agents of emerging complexity for intergalactic gas in the simulated Universe.

\begin{warning}{How can complexity analysis measure different cosmologies?}

Thanks to statistical complexity, it is possible to investigate whether even small variations on the set of cosmological parameters produce a more complex evolution of large-scale structures.  In detail, here I restrict the analysis to a few relevant variations of the baseline concordance $\Lambda$CDM model (see Fig.\ref{fig:evol2} for the detail on assumed parameters) while in \citet{va19info} I have also explored the simpler CDM case.
  
Figure \ref{fig:evol2} gives the total statistical complexity for gas residing in the  cosmic web, $T \geq 10^4 \rm K$ (normalised to a comoving $\rm Gpc^3$ volume) as a function of time for four different resimulations of the same volume considered above, for different variations of the $\sigma_8$ parameters, or the fiducial set of parameters from PLANCK \citep[][]{2016A&A...594A..13P}. 

The difference between all $\Lambda$CDM models is overall tiny  ($\leq 30\%$) at most epochs and for all energies. There clearly is a dependence between the amplitude of $\sigma_8$ (which indicates the amplitude of the initial matter power spectrum within a reference scale of $8$ comoving Mpc) and the final complexity, with a quite regular increase of complexity  going from $\sigma_8=0.7$ to $\sigma_8=0.9$. 

\begin{figure}[b]
\sidecaption
\includegraphics[width=0.6\textwidth]{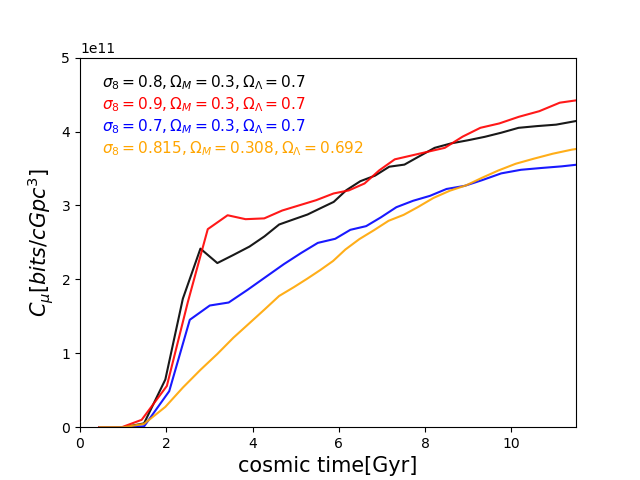}
\caption{Evolution of the total complexity in the cosmic web, for 
four different variations of cosmological parameters for the same volume.}
\label{fig:evol2}
\end{figure}  

This follows from the fact that  $\sigma_8$ is known to correlate with the rate of structure formation, and hence with the frequency of perturbations to the gas, driven by mergers. 
A higher $\sigma_8$ implies that the collapse of self-gravitating halos can begin earlier in time, and also more massive substructures within halos are present at all redshifts \citep[e.g.][]{2012ARA&A..50..353K}. In the case of a high $\sigma_8$, also more shocks are launched in the cosmic volume by the infall of gas matter \citep[see Appendix of][]{va09shocks}. All these factors make the high $\sigma_8$ Universe slightly more complex than a low $\sigma_8$ one, at all epochs.
In \citet{va19info}  I have also shown that complexity analysis can easily detect 
a large difference between $\Lambda$CDM and CDM models, even if the  CDM model is calibrated to reproduce approximately the same number of halos in the volume.  While the total number of clusters forming in the volume approximately scale as $\Omega_M \sigma_8^0.5$  \citep[e.g.][]{2002ARA&A..40..539R}, the growth of cluster is significantly delayed in the CDM cosmology \citep[e.g.][]{2001ApJ...551...15B}, due to the lower  $\sigma_8$ (0.43 in this case),  and complex
pattern driven by matter accretions in halos only emerge at later times.

\end{warning}

\section{Conclusions}

In the present epoch of Big Data in astrophysics, as a result of existing and future 
 multi-messenger surveys (as well as by increasingly more sophisticated numerical simulations), identifying the complex chain of processes which lead to observed astronomical phenomena is an open challenge, calling for new and powerful analysis approaches. \\
 
 In this contribution, I have discussed how new statistical tools based on Information Theory \citep[e.g.][]{1998PhLA..238..244F,adami,prokopenko2009information} allow us to  objectively measure the level of complexity of modern astrophysical simulations, and in the process also to identify patterns and sequences of events otherwise impossible to identify with standard approaches. \\
 
  Complexity enables us to identify which physical processes are mostly responsible for the emergence of observed complex dynamical behaviour across cosmic epochs and environments, and possibly to improve numerical  refinement strategies in future simulations attempting to reproduce the Universe. \\

 With this method I have showed that the complexity of cosmic structures has emerged early in time, when most of seeds of halos in the cosmic web started to collapse and convert their gravitational infall energy into thermal energy and magnetic field. The process is mostly mediated by violent fluid perturbations, often in the form of strong shocks and turbulent motions. On smaller scales, and before the formation of halos,  the activity connected with the formation of galaxies (e.g. radiative gas cooling and feedback from active galactic nuclei) can  introduce more complexity to the evolution of baryons in the cosmic web, which can be identified even at later simulated epochs. \\
 
 It must be noticed that the concept of complexity used in this work is a dynamical, rather than a geometrical/topological one; the latter approach has been instead  explored in works that studied cosmic structure using Minkowski functionals and Betti numbers as a proxy for topological persistence of structures \citep[e.g.][and references therein]{2017MNRAS.465.4281P}. 
 
 In passing, we can also remark that the dynamical view of complexity exposed here is different from the definition of {\it maximum} information usually adopted by the  holographic description of the Universe \citep[e.g.][]{2004ConPh..45...31B,2005bhis.book.....S}, which yields the astounding maximum information capacity of $\sim 10^{100} \rm bits$ which can be stored in the entire fabric of space-time using all available Planck lengths thereby contained   \citep[e.g.][]{2003SciAm.289b..58B}.

Based on the complexity measured in my simulation \citep[][]{va19info}, it is possible to extrapolate the total statistical  complexity within the entire observable Universe: 

\begin{eqnarray}
               C_{\rm Universe}= 4 \pi \int_0^{\infty} \langle C_{\rm \mu}(z)\rangle  r^2 \frac{dr}{dz}dz
             \approx 3.56 \cdot 10^{16} ~\rm bits \approx 4.3 ~\rm Pb.
\end{eqnarray}

where the integration is done using the redshift evolution of $C_{\rm \mu}$ measured in the fiducial $\Lambda$CDM simulation, and $dr/dz$ has been measured as a function of redshift and of the cosmological model \citep[e.g.][]{2018PASP..130g3001C}.
Interestingly, this is of the same order of the total amount of data daily generated by social media {\footnote{ http://res.cloudinary.com/yumyoshojin/image/upload/v1/pdf/future-data-2019.pdf}}, and this is also similar to the latest estimates of the maximum memory capacity of the human brain,  which follows from 
the extrapolation of the information that can be stored by synaptic plasticity
\citep[][]{ISI:000373445100001} {\footnote{See also http://nautil.us/issue/74/networks/the-strange-similarity-of-neuron-and-galaxy-networks-rp for a recent semi-quantitative comparison between the structural properties of the cosmic web and of the human neuronal network.}}.\\

 
 The above memory capacity estimate represents the minimum amount of information required to describe the evolution of the entire visible cosmic web with a spatial detail of $\sim 10^2$ kpc, which is the one that maximises the efficiency of prediction (Sec. \ref{subsubsec:eff}).
 
  This information can be crucial to define the optimal approaches for future simulations, aiming at matching the sky and redshift coverage of  incoming wide and deep multi-band surveys of the sky wavelengths (e.g. from Euclid to the Square Kilometre Array), which will define the future of astronomical data.\\  
 
 The expected flurry of complex data produced by such surveys  will keep  challenging simulators to produce Universes containing an equal amount of complexity. 
 The new analysis methods offered by Information Theory, and described in this work, promise to offer simulators with objective tools to embark on this exciting challenge that awaits in the near future.

\begin{acknowledgement}
I acknowledge financial support from the ERC  Starting Grant "MAGCOW", no. 714196
The cosmological simulations were performed with the ENZO code (http://enzo-project.org), which is the product of a collaborative effort of scientists at many universities and national laboratories..   The simulations on which this work is based have been produced on the J\"ulich Supercomputing Centre (JFZ) under project HHH42 and {\it stressicm}, as well as on Marconi at CINECA (Bologna, Italy), under project INA17\_C4A28 (in all cases with F.V. as Pricipal Investigator). 
I also acknowledge the usage of online storage tools kindly provided by the INAF Astronomical Archive (IA2) initiave (http://www.ia2.inaf.it). 
\end{acknowledgement}

\bibliographystyle{mnras}
\bibliography{franco,info}

\end{document}